\documentstyle[amsfonts,citesort]{article} 

\begin{document}  

\title{Slightly Bimetric Gravitation}

\author{J. Brian Pitts\footnote{The Ilya Prigogine Center for Studies in      Statistical Mechanics and Complex Systems      RLM 7.208      The University of Texas at Austin      Austin, TX, USA, 78712       telephone (512) 471-7253      fax (512) 471-9621      email jpitts@physics.utexas.edu  } 
and W.C. Schieve\footnote{The Ilya Prigogine Center for Studies in      Statistical Mechanics and Complex Systems      RLM 7.208      The University of Texas at Austin      Austin, TX, USA, 78712}} 
   \date{\today}  

\maketitle

  \begin{abstract}       The inclusion of a flat metric tensor in gravitation permits the formulation of a gravitational stress-energy tensor and the formal derivation of general relativity from a linear theory in flat spacetime.  Building on the works of Kraichnan and Deser, we present such a derivation using universal coupling and gauge invariance.     

    Next we slightly weaken the assumptions of universal coupling and gauge invariance, obtaining a larger ``slightly bimetric'' class of theories, in which the Euler-Lagrange equations depend only on a curved metric, matter fields, and the determinant of the flat metric.  The theories are equivalent to generally covariant theories with an arbitrary cosmological constant and an arbitrarily coupled scalar field, which can serve as an inflaton or dark matter.    	

The question of the consistency of the null cone structures of the two metrics is addressed.    \end{abstract} 

 keywords:  bimetric, causality principle, unimodular, null cone   
\section{Introduction}       A number of authors \cite{Fierz,FierzPauli,Rosen1,Weyl,Papapetrou,Gupta,Kraichnan,
Gutman,Burlankov,Thirring,Halpern,HalpernComp,Feynman,OP,Weinberg65,Sexl,NSSexlField,NSSexlLinear,Deser,Weinberg,WeinbergWitten,PenroseWeinberg,van Nieuwenhuizen,Groenewold,DeserQG,DeserFermion,Fronsdal,Cavalleri,LogunovFund,Fuchs,Gottlieb,Nikolic,Nikishov,Straumann}  have discussed the utility of a flat background metric $\eta_{\mu\nu}$ in general relativity or the possibility of deriving that theory, approximately or exactly, from a flat spacetime  theory.\footnote{For  completeness, we note that general relativity has also been derived from self-interaction on curved backgrounds \cite{Grishchuk,DeserCurved}.  Also, the utility of a background metric (in this case dynamical) in defining Lagrangian densities and conserved quantities, has recently been discussed by L. Fatibene \emph{et al.} \cite{FatFerFra}.}  Doing so enables one to formulate a gravitational stress-energy tensor \cite{Babak}, not merely a pseudotensor, so gravitational energy-momentum is localized  in a coordinate-independent way.  It also enables one to derive general relativity and other generally covariant theories, rather than merely postulating them.  (We call a theory ``generally covariant'' if no nondynamical fields appear in the Euler-Lagrange equations, even if some do appear in the action.)  As W. Thirring observed, it is not clear \emph{a priori} why Riemannian geometry is to be preferred over all the other sorts of geometry that exist, so a derivation is attractive \cite{Thirring}.  Furthermore, a non-geometrical form of gravitation can facilitate introduction of supersymmetry \cite{Zel1}.  

\section{Generally Covariant Theories from Universal Coupling and Infinitesimal Free Field Action Gauge Invariance}      To such a derivation of generally covariant theories we now turn.  Our derivation combines elements familiar from the work of Kraichnan \cite{Kraichnan} and Deser \cite{Deser}, but it has  improvements as well.  It is based upon universal coupling and an assumed initial infinitesimal invariance (up to a boundary term) of the free gravitational action.  This derivation will also serve as the model for the new derivation of slightly bimetric theories.  The assumption of gauge invariance requires that the field be massless.    
\subsection{Free Field Action}      Let $S_{f}$ be the action for a free symmetric tensor field $\gamma_{\mu\nu}$ (of density weight 0) in Minkowski spacetime with metric tensor $\eta_{\mu\nu}$ in arbitrary coordinates.  The torsion-free $\eta$-compatible covariant derivative is denoted by $\partial_{\mu}$.   The field $\gamma_{\mu\nu}$ will turn out to be the gravitational potential.  We require that $S_{f}$ change only by a boundary term under the infinitesimal gauge transformation $ \gamma_{\mu\nu} \rightarrow  \gamma_{\mu\nu}  + \delta \gamma_{\mu\nu}$, where \begin{eqnarray} \delta \gamma_{\mu\nu} = \partial_{\mu} \xi_{\nu} + \partial_{\nu} \xi_{\mu},  \label{gaugeinv} \end{eqnarray} $\xi_{\nu}$ being an arbitrary covector field.   (By changing the density weights of the fields, one could use invariance under $\delta \gamma_{\mu\nu} = \partial_{\mu} \xi_{\nu} + \partial_{\nu} \xi_{\mu} + c \eta_{\mu\nu} \partial_{\alpha} \xi^{\alpha}$ for $c \neq - \frac{1}{2}$; the case  $c = -\frac{1}{2}$ gives merely a scalar theory \cite{OP}.   The other cases, \emph{mutatis  mutandis}, give the same result as the weight 0 case.)   In the special case that the Lagrangian density is a linear combination of terms quadratic in first derivatives of the $\gamma_{\mu\nu}$, and free of algebraic and higher-derivative dependence on $\gamma_{\mu\nu}$, the requirement of gauge invariance uniquely fixes coefficients of the terms in the free field action up to a boundary term \cite{Hakim}, giving linearized vacuum general relativity \cite{Ohanian}.\footnote{For related work, one might see Wald \cite{WaldSpin} and Heiderich and Unruh \cite{UnruhHeiderich}.}

       For any $S_{f}$ invariant in this sense under (\ref{gaugeinv}), the free field equation is identically divergenceless, as we now show.  With arbitrary divergences $e^{\mu},_{\mu}$ and $f^{\mu},_{\mu}$ permitted, the action changes by  \begin{eqnarray} \delta S_{f} = \int d^{4}x  \left[\frac{\delta S_{f} }{ \delta \gamma_{\mu\nu} }  (\partial_{\nu} \xi_{\mu} + \partial_{\mu} \xi_{\nu}) + e^{\mu},_{\mu}\right]= \int d^{4}x f^{\mu},_{\mu}. \end{eqnarray} The explicit forms of the boundary terms are not needed for our purposes.  Integrating by parts, letting $\xi^{\mu}$ have compact support to annihilate the boundary terms (as we shall do throughout the paper), and making use of the arbitrariness of $\xi^{\mu}$, we obtain the identity \begin{eqnarray} \partial_{\mu} \frac{\delta S_{f}  } {  \delta \gamma_{\mu\nu} }  = 0. \label{gaugeresult} \end{eqnarray}

 \subsection{Metric Stress-Energy Tensor}      If the energy-momentum tensor is to be the source for the field $\gamma_{\mu\nu}$, consistency requires that the \emph{total} energy-momentum tensor be used, including gravitational energy-momentum, not merely nongravitational (``matter'') energy-momentum, for only the total energy-momentum tensor is divergenceless in the sense of $\partial_{\nu}$ \cite{Deser}, or, equivalently, in the sense of a Cartesian coordinate divergence.  To obtain a global conservation law, one needs a vanishing \emph{coordinate} divergence for the 4-current.  In general relativity in its geometrical form, one must choose between tensorial expressions and global conservation laws.  If one employs only tensors (or tensor densities), one can write $\nabla_{\mu} T^{\mu\nu}_{\rm mat} = 0$ for the matter stress tensor (where $\nabla_{\mu}$ is the usual torsion-free $g$-compatible covariant derivative).  But this equation typically does not yield a global conservation law \cite{Wald}, because in general it cannot be written as a  coordinate divergence.  (From the flat spacetime viewpoint, this equation is best regarded as a force law, not a conservation equation.)  If coordinate-dependent expressions are admitted, then one can write $\tau^{\mu\nu},_{\mu}=0$, where $\tau^{\mu\nu}$ is some nontensorial complex that includes gravitational as well as matter energy-momentum \cite{Adler,Stephani}.  But these objects behave oddly under coordinate transformations \cite{Pauli,Moller,MollerItaly,MollerWarsaw,MollerSurvey,RosenLocal,Anderson}.  A flat background metric, in contrast, yields \emph{tensorial global} conservation laws, as Rosen has emphasized \cite{RosenLocal,RosenWarsaw}.   Whether this stress tensor is entirely satisfactory will be considered below.  

	An expression for the total energy-momentum tensor can be derived from $S$ using the metric recipe \cite{Kraichnan,Babak,Anderson} in the following way.  The action depends on the flat metric $\eta_{\mu\nu}$, the gravitational potential $\gamma_{\mu\nu}$, and bosonic matter fields $u$.  Here $u$ represents an arbitrary collection of dynamical tensor fields of arbitrary rank, index position, and density weight.  Under an arbitrary infinitesimal coordinate transformation described by a vector field $\xi^{\mu}$, the action changes by the amount  \begin{eqnarray} \delta S = \int d^{4}x \left(\frac{\delta S}{\delta \gamma_{\mu\nu}} \pounds_{\xi} \gamma_{\mu\nu}  + \frac{\delta S}{\delta u} \pounds_{\xi} u + \frac{\delta S}{\delta\eta_{\mu\nu} } \pounds_{\xi} \eta _{\mu\nu} + g^{\mu},_{\mu} \right). \end{eqnarray}   But $S$ is a scalar, so $\delta S= 0$.  Letting the matter and gravitational field equations hold gives  \begin{eqnarray} \delta S = \int d^{4}x  \frac{\delta S}{\delta \eta_{\mu\nu} } \pounds_{\xi} \eta_{\mu\nu} = 0,  \label{conserve} \end{eqnarray} or \begin{eqnarray} \partial_{\mu} \frac{\delta S}{\delta \eta_{\mu\nu} } = 0. \end{eqnarray} This metric energy-momentum tensor density ${\frak{T}}^{\mu\nu} = 2 \frac{\delta S}{\delta \eta_{\mu\nu} } $ agrees with the symmetrized canonical tensor in the case of electromagnetism, up to a trivial factor (assuming the electromagnetic potential to be a covector of vanishing density weight, \emph{i.e.}, a 1-form; otherwise, terms that vanish when the equations of motion hold also arise).  In more general cases, the relation between the metric and symmetrized canonical results is more complicated, so some ambiguity in the term ``energy-momentum tensor'' exists; one could try to resolve this ambiguity by introducing further criteria \cite{Babak,Anderson,Hehl}.

 \subsection{Choice of Dynamical Variables} 	Deser treated the gravitational potential and $\{ ^{\mu}_{\alpha\beta} \}$ as independent variables, giving a first-order Lagrangian formalism \cite{Deser}.  This approach, which lacks Lagrange multipliers to enforce the Levi-Civita character of the connection, can be made to work if one is clever, but we prefer using only $\gamma_{\mu\nu}$ as the independent variable, as in Kraichnan's second-order Lagrangian approach \cite{Kraichnan}.  There are several reasons for our preference.  First, the second order approach seems more natural \cite{HehlVary} and physical because it avoids unnecessary variables (40 extra ones).  In Deser's derivation, the connection is just Levi-Civita's on-shell, so its dynamics is not interesting.  Second, as Deser's approach simply \emph{verifies} that an assumed from is correct, it requires either a lucky guess or knowledge of the answer in advance, whereas the second-order recipe does not.  Furthermore, the second-order approach is cleaner and more elegant, for no messy calculations are required.  Finally, this second-order approach is more general in two respects.  First, all generally covariant theories, including those with higher derivatives, manifestly fall within its scope, rather than remaining latent possibilities in the form of other lucky guesses.  Second, the first-order approach either fails if the matter action depends on the connection \cite{Ray}, as it does for a perfect fluid \cite{Lindstrom}, or requires the introduction of still more variables (perhaps another 40) to serve as Lagrange multipliers.  In contrast, the second order approach always works using only 10 variables.  For these reasons, we find a second-order principle preferrable. 

   \subsection{Full Universally-Coupled Action} 
	We seek an action $S$ obeying the physical requirement that the Euler-Lagrange equations be just the free field equations for $S_{f}$ augmented by the total energy-momentum tensor: \begin{eqnarray} \frac{\delta S}{\delta \gamma_{\mu\nu} } = \frac{\delta S_{f} }{\delta \gamma_{\mu\nu} } - \lambda \frac{\delta S}{\delta \eta_{\mu\nu} },  \label{Universal} \end{eqnarray}   where $\lambda = - \sqrt{32 \pi G}$.  In this respect our derivation follows Deser's more than Kraichnan's, for Kraichnan made no use of a free field action, but only of postulated free field equations.   

 	The basic variables in this approach are the gravitational potential $\gamma_{\mu\nu}$ and the flat metric $\eta_{\mu\nu}$.  But one is free to make a change of variables in $S$ from $\gamma_{\mu\nu}$ and $\eta_{\mu\nu}$  to $g_{\mu\nu}$ and $\eta_{\mu\nu}$, where \begin{eqnarray} g_{\mu\nu} = \eta_{\mu\nu}  - \lambda \gamma_{\mu\nu}. \end{eqnarray}  Equating coefficients of the variations gives \begin{eqnarray}  \frac{\delta S}{\delta \eta _{\mu\nu}} |\gamma =   \frac{\delta S}{\delta \eta_{\mu\nu}} |g  +  \frac{\delta S}{\delta g_{\mu\nu}}   \end{eqnarray} and  \begin{eqnarray}     \frac{\delta S}{\delta \gamma _{\mu\nu}}  =  - \lambda \frac{\delta S}{\delta g _{\mu\nu}}.  \label{ELVarChange} \end{eqnarray} Putting these two results together gives \begin{eqnarray}  \lambda \frac{\delta S}{\delta \eta_{\mu\nu}} |\gamma =  \lambda   \frac{\delta S}{\delta \eta_{\mu\nu}} |g  -  \frac{\delta S}{\delta \gamma_{\mu\nu}}.   \label{StressSplit} \end{eqnarray}  Equation (\ref{StressSplit}) splits the stress tensor into one piece that vanishes when gravity is on-shell and one piece that does not.  Using this result in (\ref{Universal}) gives  \begin{eqnarray} \lambda \frac{\delta S}{\delta \eta _{\mu\nu}} |g =  \frac{\delta S_{f}}{\delta \gamma _{\mu\nu}}, \label{Key} \end{eqnarray} which says that the free field Euler-Lagrange derivative must equal (up to a constant factor) that part of the total stress tensor that does not vanish when the gravitational field equations hold.  Recalling (\ref{gaugeresult}), one derives  \begin{eqnarray} \partial_{\mu} \frac{\delta S}{\delta \eta _{\mu\nu}} |g = 0,  \label{curl} \end{eqnarray} which says that the part of the stress tensor not proportional to the gravitational field equations has identically vanishing divergence (on either index), \emph{i.e.}, is a (symmetric) ``curl'' \cite{Anderson}. This result concerning the splitting of the stress tensor will be used in considering the gauge transformations of the full theory.  It also ensures that the gravitational field equations \emph{alone} entail conservation of energy-momentum, without any separate postulation of the matter equations.  Previously the derivation of a conserved stress tensor required that gravity \emph{and matter} obey their field  equations, as in (\ref{conserve}).  This is possible only if the gravitational potential encodes considerable information about the matter fields through constraints. The Hamiltonian and momentum constraints imply this very fact \cite{Wald}, so one sees the origin of constraints from another angle.

   We observe that the quantity $\frac{\delta S}{ \delta \eta _{\mu\nu}} |g$, being symmetrical and having identically vanishing divergence on either index, is of the form \begin{eqnarray}    \frac{\delta S}{ \delta \eta _{\mu\nu}} |g = \frac{1}{2}  \partial_{\rho} \partial_{\sigma} (  {\mathcal{M}} ^{[\mu\rho][\sigma\nu]} +   {\mathcal{M}} ^{[\nu\rho][\sigma\mu]} )  + b \sqrt{-\eta} \eta^{\mu\nu} \end{eqnarray} \cite{Wald} (pp. 89, 429), where ${\mathcal{M}} ^{\mu\rho\sigma\nu}$ is a tensor density of weight $1$ and $b$ is a constant.  This result follows from the converse of Poincar\'{e}'s lemma in Minkowski spacetime.  (It is not strictly necessary to  separate the $b$ term out, but doing so is convenient, because getting this term from ${\mathcal{M}} ^{\mu\rho\sigma\nu}$ would require that ${\mathcal{M}} ^{\mu\rho\sigma\nu}$  depend on the position 4-vector.)  We gather all dependence on $\eta_{\mu\nu}$ (with $g_{\mu\nu}$ independent) into one term, writing  \begin{eqnarray}    S = S_{1} [g_{\mu\nu}, u] + S_{2}[g_{\mu\nu}, \eta_{\mu\nu}, u]. \end{eqnarray} One easily verifies that if \cite{Kraichnan}   \begin{eqnarray}    S_{2} = \frac{1}{2} \int d^{4}x R_{\mu\nu\rho\sigma} (\eta) {\mathcal{M}} ^{\mu\nu\rho\sigma} (\eta_{\mu\nu}, g_{\mu\nu}, u ) + \int d^{4}x \alpha^{\mu},_{\mu} + 2 b \int d^{4}x \sqrt{-\eta},  \end{eqnarray} then $ \frac{\delta S_{2} }{  \delta \eta _{\mu\nu}}  |g $ has just the desired form, while $S_{2}$ does not affect the Euler-Lagrange equations.  While Kraichnan's derivation has the advantage of not needing the physical answer beforehand, it does require clever mathematical use of the flat spacetime Riemann tensor to obtain superpotential-like terms.  This quantity tends to be overlooked because it vanishes, but it is useful because its variation does not.  The boundary and 4-volume terms are novel and useful, though not essential. The boundary term is necessary for showing that Rosen's action (with no second derivatives of the dynamical variables) can be derived \emph{via} universal coupling in flat spacetime, not merely postulated.\footnote{Although such a derivation was never presented by Rosen, to our knowledge, he did indicate that such a derivation would be desirable and intended to complete the project himself \cite{Rosen1} (p. 153 of the second paper from 1940).  As he notes, deriving the theory from flat spacetime seems more appealing than merely   grafting the flat metric onto  general relativity after the fact.} The 4-volume term can cancel the 0th order term in the action, so that the action vanishes when there is no gravitational field.

   Thus,  \begin{eqnarray}    S = S_{1} [g_{\mu\nu}, u] + \frac{1}{2} \int d^{4}x R_{\mu\nu\rho\sigma} (\eta) {\mathcal{M}} ^{\mu\nu\rho\sigma} + 2 b \int d^{4}x \sqrt{-\eta}  + \int d^{4}x \partial_{\mu} \alpha^{\mu} . \end{eqnarray} The boundary term is at our disposal.  $\alpha^{\mu}$ is a weight $1$ vector density, because we require that $S$ be a scalar.  For $S_{1}$, we choose the Hilbert action for general relativity plus minimally coupled matter and a cosmological constant: \begin{eqnarray}    S_{1} = \frac{1}{16 \pi G} \int d^{4}x \sqrt{-g} R(g)  - \frac{\Lambda}{8 \pi G} \int d^{4}x \sqrt{-g} + S_{\rm mat}[g_{\mu\nu}, u]. \end{eqnarray} As is well-known, the Hilbert action is the simplest (scalar) action that can be constructed using only the metric tensor.  If the gravitational field vanishes everywhere, then the gravitational action ought to vanish also, so we set $b=\Lambda/16 \pi G$.

    Rosen \cite{Rosen1} noted that  \begin{eqnarray}    R_{\mu\nu}(g) = R_{\mu\nu}(\eta) + E_{\mu\nu}(g, \partial),  \label{RosenRicci} \end{eqnarray} where $ E_{\mu\nu}(g, \partial) $ is identical in form to the Ricci tensor for $g_{\mu\nu}$, but with $\eta$-covariant derivatives $\partial_{\mu}$ replacing partial derivatives.  Thus one finds that \begin{eqnarray}    E_{\mu\nu}(g, \partial) = \partial_{\sigma}  \Delta _{\rho\mu}^{\sigma}  - \partial_{\mu}  \Delta _{\rho\sigma}^{\sigma}  +  \Delta _{\rho\mu}^{\alpha} \Delta _{\alpha\sigma}^{\sigma} - \Delta _{\rho\sigma}^{\alpha} \Delta _{\alpha\mu}^{\sigma},      \end{eqnarray} where the field strength tensor $\Delta _{\mu\alpha}^{\beta}$ is defined by \begin{eqnarray}    \Delta _{\mu\alpha}^{\beta}  =  \{ _{\mu\alpha}^{\beta} \} - \Gamma _{\mu\alpha}^{\beta}. \label{Delta}  \end{eqnarray} Here $\{ _{\mu\alpha}^{\beta} \}$ and $\Gamma _{\mu\alpha}^{\beta}$ are  the Christoffel symbols  for $g_{\mu\nu}$ and $\eta _{\mu\nu}$, respectively.   Using (\ref{RosenRicci}) in the Hilbert term and using the product rule on the second derivatives in $ E_{\mu\nu}(g, \partial) $ leaves first derivatives of the gravitational field and a boundary term.  The boundary term is canceled if one chooses \begin{eqnarray} 16 \pi G \alpha^{\mu} = - \Delta _{\rho\sigma}^{\mu}  {\frak g}^{\sigma\rho} + \Delta _{\rho\sigma}^{\sigma}  {\frak g}^{\mu\rho}, \end{eqnarray} where ${\frak g}^{\mu\rho}$ is the contravariant metric density of weight 1.  Using another of Rosen's results concerning the bimetric formalism \cite{Rosen1}, one readily expresses the $g$-covariant derivative of a tensor density in terms of the $\eta$-covariant derivative and terms involving $\Delta _{\rho\sigma}^{\mu}$ in place of the partial derivative and terms involving $\{ _{\rho\sigma}^{\mu} \}$.  A (1,1) tensor density of weight $w$ is illustrative.  For such a field, the $\eta$-covariant derivative \cite{Israel} is \begin{eqnarray}    \partial_{\mu} \phi^{\alpha}_{\beta} = \phi^{\alpha}_{\beta},_{\mu} + \phi^{\sigma}_{\beta} \Gamma_{\sigma\mu}^{\alpha} - \phi ^{\alpha}_{\sigma} \Gamma_{\beta\mu}^{\sigma} - w \phi^{\alpha}_{\beta} \Gamma_{\sigma\mu}^{\sigma}, \end{eqnarray} and the $g$-covariant derivative $\nabla_{\mu} \phi^{\alpha}_{\beta}$ is analogous, with connection $\{ _{\sigma\mu}^{\alpha} \}$.    Recalling equation (\ref{Delta}), one writes Rosen's result as   \begin{eqnarray}    \nabla _{\mu} \phi^{\alpha}_{\beta} =  \partial  _{\mu} \phi^{\alpha}_{\beta}   + \phi^{\sigma}_{\beta} \Delta_{\sigma\mu}^{\alpha} - \phi ^{\alpha}_{\sigma} \Delta_{\beta\mu}^{\sigma} - w \phi^{\alpha}_{\beta} \Delta_{\sigma\mu}^{\sigma}. \end{eqnarray}

   The action to date takes the form \begin{eqnarray} S = \frac{1}{16 \pi G}\int d^{4}x {\frak g}^{\mu\rho} R_{\mu\rho}(\eta) \nonumber + \frac{1}{2} \int d^{4}x R_{\mu\nu\rho\sigma} (\eta) {\mathcal{M}} ^{\mu\nu\rho\sigma} (\eta_{\mu\nu}, g_{\mu\nu}, u )\nonumber \\ + \frac{1}{16 \pi G} \int d^{4}x {\frak g}^{\mu\rho} (\Delta _{\mu\alpha}^{\sigma} \Delta _{\rho\sigma}^{\alpha} - \Delta _{\rho\mu}^{\sigma} \Delta _{\alpha\sigma}^{\alpha}) + S_{\rm mat}[g_{\mu\nu},u].  \end{eqnarray} One can make $R_{\mu\nu\rho\sigma} (\eta)$ disappear from $S$ by setting  \begin{eqnarray} {\mathcal{M}} ^{\mu\nu\rho\sigma} = -  \eta^{\nu\sigma} {\frak g} ^{\mu\rho}/8 \pi G. \end{eqnarray} The contravariant weight 1 metric density ${\frak g}^{\mu\rho}$ distinguishes itself here.  This quantity has often appeared to be the preferred variable, not only in flat  spacetime forms of general relativity (\emph{e.g.}, \cite{Papapetrou,Gupta}), but also in other contexts.  The DeDonder gauge condition, also known as the harmonic coordinate condition, prefers this variable \cite{Fock,KucharHarmonic}; the desirability of this gauge was strongly urged by Fock.  More recently, A. Anderson and J. York have found the  ``slicing density'' \cite{YorkHamilton}, a weight $-1$ densitized version of the ADM lapse, to be quite useful.  The slicing density is simply related to the 0-0 component of ${\frak g} ^{\mu\nu}$ \cite{YorkConference}.  One reason that we do not use ${\frak g} ^{\mu\nu}$ (or rather, $\frac{  {\frak g}^{\mu\nu} - \sqrt{-\eta} \eta^{\mu\nu} }{ \lambda }$) as the gravitational potential is to make clear that no preference for this variable is built in by hand.

     The total action is therefore Rosen's tensorial one with no second derivatives: \begin{eqnarray}  S   =  \frac{1}{16 \pi G}\int d^{4}x {\frak g}^{\mu\rho} (\Delta_{\mu\alpha}^{\sigma} \Delta_{\rho\sigma}^{\alpha} - \Delta_{\rho\mu}^{\sigma} \Delta_{\alpha\sigma}^{\alpha}  )  + S_{\rm mat}[g_{\mu\nu}, u]. \end{eqnarray} This action should be compared to those available in geometrical general relativity, where one chooses either to include second derivatives of the dynamical variables, or to give up the scalar character of the action.     

     Babak and Grishchuk \cite{Babak} have proposed a different principle for specifying ${\mathcal{M}} ^{\mu\nu\rho\sigma}$, with different results.  Their proposal gives the most desirable form to the metric stress tensor, \emph{viz.}, a tensorial relative of the Landau-Lifshitz pseudotensor \cite{Landau}, which is the only symmetric pseudotensor with no second derivatives.  This tensor had been previously obtained in a conservation law for bimetric general relativity by Rosen \cite{Rosen1}, but that derivation did not involve Noether's theorem \cite{Katz}.

  There are two key ingredients in the derivation of generally covariant theories in this way.  One is universal coupling, which says that the source for the field equations must be the total stress-energy tensor.  The other key ingredient can be either free field gauge invariance of the assumed form or gravitation-induced conservation of energy-momentum.  Gauge invariance might be motivated, if in no other way, by a desire for Lorentz invariance and positive energy.  However, as unimodular general relativity and the theories with dynamical $\frac{ \sqrt{-g} }{ \sqrt{-\eta} }$ below show, this specific form of gauge invariance more restrictive than necessary for positive energy and Lorentz invariance.  This fact follows from the fact that slightly bimetric theories behave like scalar-tensor theories (as will be shown below), and at least some of the latter have positive energy \cite{Santiago}.  This condition is therefore weaker than that required by Fierz \cite{Fierz} and van Nieuwenhuizen \cite{van Nieuwenhuizen}, who were interested in good behavior of \emph{free} fields.  One might also see G. Cavalleri and G. Spinelli \cite{Cavalleri}.  

 \subsection{Gauge Invariance and Gauge Fixing}      It is instructive to determine what has become of the original free field gauge invariance.  The scalar character of the action entails \begin{eqnarray} \delta S_{coord} = \int d^{4}x \left[\frac{\delta S}{\delta g_{\mu\nu}} \pounds_{\xi} g_{\mu\nu}  + \frac{\delta S}{\delta u} \pounds_{\xi} u + \left(\frac{\delta S}{\delta\eta_{\mu\nu} }|g \right)\pounds_{\xi} \eta _{\mu\nu} + h^{\mu},_{\mu} \right] = 0 \end{eqnarray}  under a coordinate transformation, where the form of $h^{\mu},_{\mu}$ is not important.  (The same will hold for the other boundary terms below.)  But in a flat spacetime theory, invariance under coordinate transformations is trivial.  A \emph{gauge} transformation, on the other hand, would be a transformation that changes the action changes only by a boundary term, but is not a coordinate transformation.  Using the coordinate transformation formula and noting that the terms involving the absolute objects do not contribute more than a divergence, one easily verifies that a (pure) gauge transformation is given by $\delta g^{\mu\nu} = \pounds_{\xi} g^{\mu\nu}$, $\delta u = \pounds_{\xi} u$, $\delta\eta^{\mu\nu}= 0,$ with $\xi^{\mu}$ arbitrary.  In showing that the term for the flat metric does not contribute nontrivially, one must recall from (\ref{curl}) above that \begin{eqnarray} \partial^{\mu} \frac{\delta S}{\delta \eta^{\mu\nu}}|g = 0 \end{eqnarray} identically.  (See also (\cite{LogunovFund}), but we do not impose any gauge condition \emph{a priori} as Logunov \emph{et al.} do.  If all the field equations should be derivable from an action, then Logunov \emph{et al.} would need to modify the way that the gauge condition arises in their work, which is by \emph{fiat}, or else restrict the values of the dynamical variables, with possible consequences for the field equations.)  Thus,  \begin{eqnarray} \delta S_{gauge} = \delta S_{coord} - \int d^{4}x \left[\left(\frac{\delta S}{\delta\eta_{\mu\nu} }|g\right) \pounds_{\xi} \eta _{\mu\nu} + i^{\mu},_{\mu} \right] = \nonumber \\ 0 - \int d^{4}x \left(- 2 \xi^{\alpha} \eta_{\alpha\mu} \partial_{\nu} \frac{\delta S}{\delta\eta_{\mu\nu} }|g + j^{\mu},_{\mu} \right). \end{eqnarray} Recalling from (\ref{curl}) above that \begin{eqnarray} \partial_{\mu} \frac{\delta S}{\delta \eta _{\mu\nu}} |g = 0 \end{eqnarray} identically, one sees that $\delta S_{gauge} $ is indeed merely a boundary term, so our guessed form of the gauge invariance is verified.  In this case, gauge transformations change (bosonic) dynamical fields in the same way that coordinate transformations do, but leave the nondynamical object $\eta_{\mu\nu}$ unchanged. If one performs simultaneously a gauge transformation and a coordinate transformation in the `opposite direction,' then the dynamical variables are unchanged, but the absolute object $\eta_{\mu\nu}$ is altered.    

      Given that coordinate-independent localization of gravitational energy-momentum is one of the attractive features of the bimetric approach to general relativity, does a gravitational stress-energy tensor fully satisfy the intuitive desire for localization?  As Zel'dovich and Grishchuk note, the arbitrariness in the pseudotensors of the geometrical variant is not eliminated by introducing a flat background metric, but merely transformed into the gauge-variance of the gravitational stress-energy tensor \cite{Zel1,Grishchuk90}.  

    As will appear below, attempting to find harmony between the null cone structures of the two metrics will require fixing the gauge, at least in part.  Doing so in a principled way will require further study.  But it is appropriate to comment briefly on possible gauge conditions.  Because they are tensorial, these conditions do not fix the coordinate system, but rather relate the flat and curved metrics.  Rosen suggests a tensorial relative of the DeDonder conditions,  \begin{eqnarray} \partial_{\nu} {\frak{g}}^{\mu\nu} = 0, \end{eqnarray} as one option \cite{Rosen1,RosenConf}.  This choice is the one imposed by Logunov \emph{et al.} \cite{LogunovFund}.  It has the attractive feature that when the coordinate system is Cartesian for the flat metric, it is harmonic (as DeDonder and Fock \cite{Fock} preferred) for the curved metric.   Another option noted by Rosen \cite{Rosen1} is  \begin{eqnarray} \kappa \equiv \sqrt{\frac{-g}{-\eta} } = 1, \end{eqnarray} \begin{eqnarray}  \partial_{[\beta} ({\frak{g}}_{\alpha ] \mu}  \partial_{\nu} {\frak{g}}^{\mu\nu} ) = 0,  \end{eqnarray} nontensorial relatives of which have been employed by Dragon, Kreuzer, and Buchm{\"{u}}ller \cite{BuDragonGaugeCosm,DragonKreuzer}.  It would make sense to fix the gauge in a way that harmonizes the two null cone structures, if possible; we know of no standard gauge conditions that achieve this goal.  Another option, if the traditional negative-energy objections to massive gravity \cite{DeserMass} can be overcome, would be to add a mass term.  M.  Visser has recently suggested that these problems in fact can be overcome \cite{Visser}.  Finding a mass term that ensures proper light cone behavior, if one even exists, would be a nontrivial task.    

\section{Slightly Bimetric Theories from Traceless Universal Coupling and Restricted Free Field Invariance}       The possibility of deriving general relativity in flat spacetime is by now well-known, though we believe the above derivation to be especially clear.  One naturally asks, can anything new, something besides general relativity and other generally covariant theories (with higher derivatives), be obtained from a procedure along these lines?  In fact, other theories can be derived.  We will now show a larger family of theories that can be obtained by making two modifications.  One relaxes universal coupling to apply only to the traceless part of the stress tensor, while the other restricts the free field gauge invariance to divergenceless vector fields.      

 Under conformal transformations, a metric tensor factors into two pieces.  One is the conformally invariant part, the densitized metric  $\tilde{ \eta}_{\mu\nu} $ of weight $-\frac{1}{2}$, which has determinant $ \tilde{ \eta  } = -1$.  This quantity determines the flat metric's null cone structure.  Its inverse, the weight $\frac{1}{2}$ density $\tilde{ \eta}^{\mu\nu} $, also has determinant $-1$. Using the matrix relation $\delta det(A) = (detA) Tr(A^{-1} \delta A)$, one sees that $\delta \tilde{ \eta}_{\mu\nu} $ and consequently $\frac{\delta S }{\delta  \tilde{ \eta}_{\mu\nu}}$ are  traceless.  The other, conformally variant factor is $\sqrt{-\eta }^{\frac{1}{2} }$, where $\eta$ is the determinant  of $\eta_{\mu\nu}$.  (We shall work with $\sqrt{-\eta }$ rather than its square root, but nothing important depends on this choice.)  Recalling the derivation of the metric stress tensor above, one sees that (apart from trivial factors) the traceless part of the stress tensor comes from $\tilde{ \eta}_{\mu\nu}$ and the trace comes from $\sqrt{-\eta }$.  As was just shown, universal coupling to the total stress tensor yields an effectively Riemannian theory.  It is known that in massless scalar gravity, universal coupling to the trace of the stress tensor yields a conformally flat Riemannian theory:  the determinant of the flat metric is completely ``clothed'' by the gravitational field \cite{Kraichnan,FreundNambu,DeserHalpern}.  Thus, one suspects that treating the traceless and trace parts of the stress tensor differently might yield interesting results.  Anticipating some of our results, we observe the pattern that whatever part of the stress tensor (the whole, the trace, or the traceless part) is universally coupled to gravity, the corresponding part of the flat metric (the whole, the determinant, or the conformally invariant part, respectively) is entirely ``clothed'' by the gravitational field and rendered unobservable (if the field is massless).

     We therefore write a general action for a gravitational field and bosonic matter as $S[ \tilde{ \eta}_{\mu\nu} , \sqrt{-\eta }, \tilde{ \gamma}_{\mu\nu}, u]$, with the gravitational field $\tilde{ \gamma}_{\mu\nu} $ taken as a density of weight $-\frac{1}{2}$ to match $\tilde{ \eta}_{\mu\nu} $.  The Lie derivative of tensor densities requires care.  For a (1, 1) density of weight $w$, the form is \cite{Israel} \begin{eqnarray} \pounds_{\xi} \phi^{\alpha}_{\beta} = \xi^{\mu} \phi ^{\alpha}_{\beta},_{\mu} - \phi^{\mu}_{\beta} \xi^{\alpha},_{\mu} + \phi^{\alpha}_{\mu} \xi^{\mu},_{\beta} + w \phi^{\alpha}_{\beta} \xi^{\mu},_{\mu}. \end{eqnarray} The form for any tensor density is readily generalized from this expression.

  The metric stress tensor can be split up into traceless and trace parts by reworking the earlier derivation.  One has   \begin{eqnarray} \delta S = \int d^{4}x \left(\frac{\delta S}{\delta \gamma_{\mu\nu}} \pounds_{\xi} \gamma_{\mu\nu}  + \frac{\delta S}{\delta u} \pounds_{\xi} u + \frac{\delta S}{\delta  \tilde{ \eta}_{\mu\nu}  } \pounds_{\xi}   \tilde{ \eta}_{\mu\nu}  + \frac{\delta S}{\delta \sqrt{- \eta } } \pounds_{\xi} \sqrt{ -\eta } \right) = 0. \end{eqnarray}   Letting the matter and gravitational field equations hold gives  \begin{eqnarray} \delta S = \int d^{4}x \left(   \frac{\delta S}{\delta  \tilde{ \eta}_{\mu\nu}  } \pounds_{\xi}   \tilde{ \eta}_{\mu\nu}  + \frac{\delta S}{\delta \sqrt{- \eta } } \pounds_{\xi} \sqrt{ -\eta } \right) =0. \end{eqnarray} Local energy-momentum conservation takes the form   \begin{eqnarray} \partial_{\mu} \left( 2 \frac{\delta S}{\delta  \tilde{ \eta}_{\mu\nu} }                     +  \frac{\delta S}{\delta \sqrt{- \eta } } \sqrt{ -\eta }  \tilde{\eta}^{\mu\nu} \right) = 0. \end{eqnarray}      It is convenient to introduce the following change of variables: \begin{eqnarray} S[ \tilde{ \eta}_{\mu\nu} , \sqrt{-\eta }, \tilde{ \gamma}_{\mu\nu} , u] = S[ \tilde{ \eta}_{\mu\nu} , \sqrt{-\eta }, \tilde{ g}_{\mu\nu} , u], \end{eqnarray} where  \begin{eqnarray}  \tilde{ g}_{\mu\nu}  = \tilde{ \eta}_{\mu\nu} - \lambda \tilde{ \gamma}_{\mu\nu}. \end{eqnarray} The reason for taking the gravitational field to be (0,2) weight $-\frac{1}{2}$ is now clear:  doing so makes it easy to add the gravitational potential to the conformally invariant part of the flat metric.  (Plainly a (2,0) weight $\frac{1}{2}$ field would work equally well, \emph{mutatis mutandis}.) Taking care with the trace, one finds that  \begin{eqnarray} \frac{\delta S}{\delta  \tilde{ \eta}_{\mu\nu}  } |\gamma = \frac{\delta S}{\delta  \tilde{ \eta}_{\mu\nu}} |g  +  \frac{\delta S}{\delta     \tilde{ g}_{\alpha\beta}} P^{\mu\nu}_{\alpha\beta} \end{eqnarray} and  \begin{eqnarray}     \frac{\delta S}{\delta \tilde{ \gamma}_{\mu\nu} }  =  - \lambda \frac{\delta S}{\tilde{ g}_{\mu\nu} } , \label{ELVarChangeSB} \end{eqnarray} where     \begin{eqnarray} P^{\mu\nu}_{\alpha\beta} = \delta^{(\mu}_{\alpha} \delta^{\nu)}_{\beta}  - \frac{1}{4} \eta^{\mu\nu} \eta_{\alpha\beta} \end{eqnarray} is the traceless symmetric projection tensor with respect to $\eta_{\mu\nu}$.  Combining these two results gives \begin{eqnarray}  \lambda \frac{\delta S}{\delta  \tilde{ \eta}_{\mu\nu} } |\gamma = \lambda \frac{\delta S}{\delta  \tilde{ \eta}_{\mu\nu} } |g  -  \frac{\delta S}{\delta      \tilde{ \gamma}_{\alpha\beta} }  P^{\mu\nu}_{\alpha\beta}, \label{StressSplitSB} \end{eqnarray} which splits the traceless part of the stress tensor into a part that vanishes on-shell and another that depends on how much of the conformally invariant part of the flat metric remains after the change of variables.

    We now introduce the physical postulate of traceless universal coupling:   \begin{eqnarray} \frac{\delta S}{\delta   \tilde{ \gamma}_{\alpha\beta}  }  P^{\mu\nu}_{\alpha\beta} = \frac{\delta S_{f} }{\delta      \tilde{ \gamma}_{\alpha\beta} }  P^{\mu\nu}_{\alpha\beta} - \lambda \frac{\delta S}{\delta  \tilde{ \eta}_{\mu\nu} } |\gamma ; \label{UniversalSB} \end{eqnarray} in words, the traceless part of the full field equations equals the traceless part of the free field equations coupled to the traceless part of the stress tensor.  This postulate will let us explore what theories, besides Riemannian and conformally flat Riemannian theories, can be obtained from a slightly relaxed version of universal coupling.  Combining equations (\ref{StressSplitSB}) and (\ref{UniversalSB}) gives  \begin{eqnarray} \lambda \frac{\delta S}{\delta  \tilde{ \eta}_{\mu\nu} } |g     =  \frac{\delta S_{f} }{\delta   \tilde{ \gamma}_{\alpha\beta} }  P^{\mu\nu}_{\alpha\beta}. \label{KeySB} \end{eqnarray} The traceless part of the free field equations must equal a term derived from how the flat metric remains in the action after the change to the bimetric variables.

    This result suggests that it would be useful to have a result concerning $\partial_{\mu} \frac{\delta S}{\delta   \tilde{ \gamma}_{\alpha\beta} }  P^{\mu\nu}_{\alpha\beta}$ derived from an infinitesimal invariance.   In order that only the traceless part of the free field equations be involved, the variation of the gravitational field ought itself to be traceless. We require that $S_{f}$ change at most by a boundary term under the infinitesimal transformation $\gamma_{\mu\nu} \rightarrow  \gamma_{\mu\nu}  + \delta \gamma_{\mu\nu}$, where $\delta \gamma_{\mu\nu} = \partial_{\mu} \xi_{\nu} + \partial_{\nu} \xi_{\mu}$, but with $\xi_{\mu}$ restricted so that  \begin{eqnarray} \partial _{\mu}  \xi ^{\mu} = 0. \end{eqnarray} Now $\xi_{\nu}$ is a density of weight $-\frac{1}{2}$.  Others using a similarly restricted invariance have restricted $\gamma_{\mu\nu} \eta^{\mu\nu}$ to vanish\cite{vanDam,UnruhUGR,Teitelboim,Dragon,Sorkin}, but we leave it arbitrary, anticipating that another degree of freedom might appear.  This gauge invariance is consistent with a non-zero mass and self-interaction potential for the trace part of the gravitational field.  Given the various reasons for which scalar fields are presently postulated, such as inflation and dark matter, it would be welcome to find an extra scalar field without postulating it \emph{ad hoc}.  (We should mention that string/membrane theory is another approach that gives a scalar field naturally.)  One can write  \begin{eqnarray} \xi^{\mu} = \partial_\nu \mathcal{F}^{\mu\nu}, \end{eqnarray} with $ \mathcal{F}^{\mu\nu}$ an arbitrary antisymmetric field of suitable weight.  Repeated integration by parts and the arbitrariness of $\mathcal{F}^{\mu\nu}$ entail that  \begin{eqnarray}  \partial_{\mu}\partial ^{[\rho }  P^{\nu]\mu}_{\alpha\beta} \frac{\delta S_{f}  }{  \delta \tilde{ \gamma}_{\alpha\beta} }   = 0 , \end{eqnarray} which means that the divergence of the traceless part of the free field equations equals the gradient of some function.  Recalling equation (\ref{KeySB}), one shows that $\partial_{\mu} \frac{\delta S}{\delta  \tilde{ \eta}_{\mu\nu} } |g  $ is a gradient. If one splits the full action $S$ into $S_{1}$ and $S_{2}$, then $S_{2}$ can take the same form as above for general relativity.  $S_{1}$ can have the form $S_{1}[ \tilde{ g}_{\mu\nu} ,\sqrt{-\eta },  u]$, with the $\tilde{ \eta}_{\mu\nu}$ absent.  We have not found any other solutions to equation (\ref{KeySB}).

   It is useful to make a further change of variables from a densitized curved metric to an ordinary one by \begin{eqnarray} g_{\mu\nu} = \tilde{ g}_{\mu\nu}  \sqrt{-\eta}^{ \frac{1}{2} }. \end{eqnarray}  The Euler-Lagrange equations change trivially:  $ \frac{\delta S }{\delta \tilde{ g}_{\mu\nu} } =  \frac{\delta S }{\delta  g_{\mu\nu} } \sqrt{-\eta}^{ \frac{1}{2} }$.  We conclude that the general action is  \begin{eqnarray} S = S_{1}[ g_{\mu\nu}, \sqrt{-\eta },  u] +   \frac{1}{2} \int d^{4}x R_{\mu\nu\rho\sigma} (\eta_{\mu\nu}) {\mathcal{M}} ^{\mu\nu\rho\sigma} (\eta_{\mu\nu}, g_{\mu\nu}, u ) + \int d^{4}x (\partial_{\mu} \alpha^{\mu}  + 2 b \sqrt{-\eta}). \end{eqnarray} We call this form ``slightly bimetric'': ``slightly'' because only the determinant of $\eta_{\mu\nu}$ enters the Euler-Lagrange equations essentially, not the whole flat metric, and ``bimetric'' because the whole of $\eta_{\mu\nu}$ is present somewhere in the theory, \emph{viz.}, in the action, in the definition of the stress tensor, and in the definition of ideal lengths and times for objects unaffected by gravity (of which there are none). The restriction of the initial invariance has the consequence that the gravitational field equations alone no longer suffice to yield conservation of energy-momentum; the matter fields $u$ must also obey their equations of motion, at least in part.  This  last result bears a resemblance to the result of Lee \emph{et. al.} \cite{LLN} that the ``matter response equations'' $\nabla_{\mu} T^{\mu\nu}_{\rm mat} = 0$ follow from the gravitational field equations if and only if no absolute objects are present in the field equations.  These equations still follow, of course, from the matter field equations, assuming that matter couples only to a curved metric \cite{Wald}.  


   We now turn to consider the gauge invariance of slightly bimetric theories.  Going through the same procedure as for generally covariant theories, we guess that a gauge transformation is given by $\delta g_{\mu\nu} = \pounds_{\xi} g_{\mu\nu}$, $\delta u = \pounds_{\xi} u$, $\delta\eta_{\mu\nu}= 0 $, but with $\xi^{\mu}$ obeying some restriction.  Here  $\xi^{\mu}$ has vanishing weight.  Thus,  \begin{eqnarray} \delta S_{gauge} = \delta S_{coord} - \int d^{4}x \left(\frac{\delta S}{\delta\eta_{\mu\nu} }|g \pounds_{\xi} \eta _{\mu\nu} + i^{\mu},_{\mu} \right) \nonumber \\ = 0 - \int d^{4}x \left(- 2 \xi^{\alpha} \eta_{\alpha\mu} \partial_{\nu} \frac{\delta S}{\delta\eta_{\mu\nu} }|g + j^{\mu},_{\mu} \right). \end{eqnarray} Recalling that \begin{eqnarray} \partial_{\mu} \frac{\delta S}{\delta \eta _{\mu\nu}} |g = \partial ^{\nu} \psi \end{eqnarray} for some scalar density $\psi$, one sees that $\delta S_{gauge} $ is indeed a boundary term if and only if $\partial_{\mu} \xi^{\mu} = 0$ (unless $\psi$ vanishes, in which case the theory is really generally covariant).  Thus, our assumed form of the invariance is verified, and the restriction on $\xi^{\mu}$ is known. The same restriction holds for the full nonlinear theory as held for the linear theory.  In this slightly bimetric case, gauge transformations change (bosonic) dynamical fields in the same way that $\eta_{\mu\nu}$-volume-preserving coordinate transformations do, but leave the absolute object $\eta_{\mu\nu}$ unchanged.

\section{Slightly Bimetric Theories Are Equivalent to Generally Covariant Theories plus a Scalar Field}    

   Having proposed the addition of a flat background metric to general relativity and noted the possibility of constructing alternative theories with this extra ingredient, Rosen himself subsequently devoted considerable energy to a particular bimetric theory of gravity (e.g., \cite {Rosen2}), hoping to avoid singularities, which afflict general relativity, and to give simpler partial differential equations for the Euler-Lagrange equations.  Although Rosen's theory passes a considerable number of empirical tests, it has difficulty with the binary pulsar \cite{Will}.  More generally, theories into which the flat metric enters the action  nontrivially will display various effects which can be tested against experiment.  Concerning the matter action, experiment strongly restricts how the flat metric can enter \cite{Will}, so it makes sense to let matter see only a curved metric, with the unclothed conformally invariant part of the flat metric absent, apart from a term containing the flat metric's Riemann tensor; such a term merely alters the stress tensor by a curl, and does not affect the field equations.  (But see \cite{Lohiya,Fabris,Bartolo} for recent interest in nonminimal coupling to scalar fields.  The assumption of  minimal coupling will not be used.)  Requiring that the matter stress tensor appear on the right side of the gravitational Euler-Lagrange equations substantially imposes the same condition \cite{Deser}.  The gravitational action has more room for a flat metric to enter, but one expects that theories with more exposed background geometry will have more trouble agreeing with experiment. If only the determinant of the flat metric $\sqrt{-\eta}$ appears in the action nontrivially, then the effects should be testable, but not as constrained as if the whole metric appears.  Slightly bimetric theories therefore are perhaps the best chance for empirically viable continuation of Rosen's bimetric program.  However, they do not satisfy Rosen's desire for simpler partial differential equations.  Whether slightly bimetric theories help to avoid singularities is tied to the success of scalar-tensor theories in doing the same.  On this point, reports are mixed \cite{Lohiya,Kaloper}.

    It is convenient to split the action into effectual and ineffectual pieces, so we write  \begin{eqnarray}     S = S_{e}[g_{\mu \nu}, \sqrt{-\eta}, u] + S_{i}[g_{\mu \nu}, \eta_{\mu\nu} u],  \end{eqnarray} both terms being scalars.  The effectual terms are those that affect the Euler-Lagrange equations.  All terms that do not affect the (gravitational or matter) Euler-Lagrange equations and that contribute at most a curl to $\frac{\delta S}{\eta_{\mu\nu} }$, \emph{viz.}, divergences, flat space $4$-volume terms, and terms involving $R_{\mu\nu\rho\sigma} (\eta_{\mu\nu})$, are gathered into the ineffectual term $S_{i}$.          

  Making use of the properties of the action under coordinate transformations, one can derive generalized Bianchi identities \cite{LLN}.  Under an arbitrary infinitesimal coordinate transformation described by a vector field $\xi^{\mu}$, the action changes by the amount   \begin{eqnarray} \delta S = \int d^{4}x \left( \frac{\delta S_{e} }{\delta g_{\mu\nu}} \pounds_{\xi} g_{\mu\nu}  + \frac{\delta S_{i} }{\delta g_{\mu\nu}} \pounds_{\xi} g_{\mu\nu}  +   \frac{\delta S_{e} }{\delta \sqrt{-  \eta } } \pounds_{\xi} \sqrt{-\eta} \right. \nonumber  \\ \left. + \frac{\delta S_{i} }{\delta \eta_{\mu\nu}}  \pounds_{\xi}  \eta_{\mu\nu}    + \frac{\delta S_{e}}{\delta u} \pounds_{\xi} u   + \frac{\delta S_{i} }{\delta u} \pounds_{\xi} u  \right) = 0. \end{eqnarray} By construction $\frac{\delta S_{i} }{\delta g_{\mu\nu}}$ and $\frac{\delta S_{i} }{\delta u}$  vanish identically, so the second and sixth terms do not contribute.   One observes that $\frac{\delta S_{i} }{\delta \eta_{\mu\nu}}$ is a curl, so the fourth term contributes only a boundary term.  Letting the matter field equations $\frac{\delta S}{\delta u} = 0$ and the gravitational field equations $\frac{\delta S}{\delta g_{\mu\nu}} = 0$ hold annihilates the first and third terms, so only the second remains: \begin{eqnarray} \delta S = \int d^{4}x   \frac{\delta S_{e}}{\delta \sqrt{-\eta}} \pounds_{\xi} \sqrt{-\eta} = 0. \end{eqnarray} Thus, upon integration, one obtains \begin{eqnarray} \frac{\delta S_{e}}{\delta \sqrt{-\eta} } = J, \end{eqnarray} where $J$ is a constant of integration.

  This last equation is sufficiently similar in appearance to an Euler-Lagrange equation that one can consider another theory with a dynamical metric, matter fields, and a \emph{dynamical} weight 1 density $\psi$, with $\psi$ replacing $\sqrt{-\eta}$, plus an additional term:    \begin{eqnarray} S'[g_{\mu \nu}, u, \psi] = S_{e}[g_{\mu \nu}, u, \psi] - \int d^{4}x J \psi. \end{eqnarray} The Euler-Lagrange equations for this action are  \begin{eqnarray} \frac{\delta S'}{\delta g_{\mu\nu}} = 0, \end{eqnarray} \begin{eqnarray} \frac{\delta S'}{\delta u} = 0, \end{eqnarray} and  \begin{eqnarray} \frac{\delta S'}{\delta \psi} = 0. \end{eqnarray} The metric and matter equations are identical to those for the original action $S_{e}$.  The equation for $\psi$ is equivalent to the integrated on-shell identity $ \frac{\delta S_{e}}{\delta \sqrt{-\eta} } = J$ above.  The theories differ in substance, for one has an absolute (\emph{i.e.}, nondynamical) object, and $J$ is an integration constant, while the other has no absolute objects, and $J$ is a parameter in the action.  But they do not differ in the forms and solutions of the equations:  they are empirically indistinguishable.  Thus, scalar density-tensor theories are equivalent to slightly bimetric theories in this sense.  We emphasize that the coupling of the scalar field to the curved metric is of arbitrary form, not necessarily minimal.

  Ordinarily one considers theories with a scalar field, not a scalar density field, so it is now useful to show that the scalar density-tensor theories above can be recast as theories with a scalar field.  This recasting involves a change of variables $\phi = \psi/\sqrt{-g}$.  Reexpressing the action $S'$ in the new variables gives  \begin{eqnarray} S''[g_{\mu \nu}, u, \phi] = S'[g_{\mu \nu}, u, \psi]. \end{eqnarray} The matter field equations are untouched by this transformation.  Under a variation of $g_{\mu\nu}$ and $\psi$, one obtains  \begin{eqnarray} \frac{\delta S'}{\delta g_{\mu\nu}} \delta g_{\mu\nu}  + \frac{\delta S'}{\delta \psi}  \delta \psi = \frac{\delta S''}{\delta g_{\mu\nu}} \delta g_{\mu\nu}  + \frac{\delta S''}{\delta \phi}  \delta \phi. \end{eqnarray} Using $\psi= \phi \sqrt{-g}$ and equating coefficients of $\delta g$ and of $\delta \phi$ gives   \begin{eqnarray} \frac{\delta S'}{\delta g_{\mu\nu }} |\psi +  \frac{\delta S'}{\delta \psi} g^{\mu\nu} \phi \sqrt{-g}/2 = \frac{\delta S''}{\delta g_{\mu\nu}} |\phi \end{eqnarray} and  \begin{eqnarray} \frac{\delta S'}{\delta \psi} \sqrt{-g} = \frac{\delta S''}{\delta \phi}. \end{eqnarray} One sees that the scalar-tensor equations are just linear combinations of the scalar density-tensor equations.  Thus every slightly bimetric theory has a scalar-tensor ``twin'' and \emph{vice versa}.      

\section{General Form for a Slightly Bimetric Theory}                                                                                   If one prohibits derivatives higher than second order (and permits those only linearly) in the Lagrangian density, then the most general slightly bimetric action is of the form \begin{eqnarray} S = \frac{1}{16 \pi G} \int d^{4}x [ a(\kappa) \sqrt{-g} R(g)  + f(\kappa) \sqrt{-g} g^{\mu\nu}\Delta_{\mu\alpha}^{\alpha} \Delta _{\nu\beta}^{\beta}  + e(\kappa)\sqrt{-g} ] \nonumber \\  +   \frac{1}{2} \int d^{4}x R_{\mu\nu\rho\sigma} (\eta_{\mu\nu}) {\mathcal{M}} ^{\mu\nu\rho\sigma} (\eta_{\mu\nu}, g_{\mu\nu}, u )  + \int d^{4}x \partial_{\mu} \alpha^{\mu}  + S_{\rm mat}[ g_{\mu\nu}, \sqrt{-\eta },  u] . \label{SBAction} \end{eqnarray} The term $2 b \sqrt{-\eta}$ has been absorbed into $e(\kappa)\sqrt{-g}$, while the possible term $c(\kappa) \sqrt{-g} g^{\mu\nu} \nabla_{\mu} \nabla_{\nu} \kappa $ has been absorbed by redefinition of $f(\kappa)$ and $\alpha^{\mu}$.  Employing Rosen's results as above, one can rewrite this action in a Rosen-esque form with no second derivatives of either dynamical or absolute variables: \begin{eqnarray}  S   =  \frac{1}{16 \pi G} \int d^{4}x {\frak g}^{\mu\rho} \left[ a\left(\kappa \right) \Delta_{\mu\alpha}^{\sigma} \Delta_{\rho\sigma}^{\alpha} - \left(a + \kappa \frac{da}{d\kappa}\right) \Delta_{\rho\mu}^{\sigma} \Delta_{\alpha\sigma}^{\alpha}  + \right. \nonumber \\ \left. \left(f(\kappa) + \kappa \frac{da}{d\kappa}\right) \Delta_{\rho\sigma}^{\sigma} \Delta_{\mu\alpha}^{\alpha}  
 +  e(\kappa) \sqrt{-g}\right]  + S_{\rm mat}[g_{\mu\nu}, \sqrt{-\eta}, u]. \end{eqnarray} In writing this form, we have set  \begin{eqnarray} 16 \pi G \alpha^{\mu} = - a(\kappa) \Delta_{\rho\sigma}^{\mu}  {\frak g}^{\sigma\rho} + a(\kappa) \Delta _{\rho\sigma}^{\sigma}  {\frak g}^{\mu\rho} \end{eqnarray} and  \begin{eqnarray} {\mathcal{M}} ^{\mu\nu\rho\sigma} = -  a(\kappa) \eta^{\nu\sigma} {\frak g} ^{\mu\rho}/ 8 \pi G. \end{eqnarray}

     Using (\ref{SBAction}) one finds the Euler-Lagrange equations of motion to be    \begin{eqnarray}    \frac{16 \pi G}{\sqrt{-g} } \frac{\delta S}{\delta g_{\mu\nu}}  = -a G^{\mu\nu}+ \frac{\kappa}{2} a' g^{\mu\nu} R +  \nabla^{\mu} \nabla^{\nu}a -  g^{\mu\nu} \nabla^{2} a  \nonumber \\ -\nabla_{\sigma}(f  g^{\mu\nu} \Delta_{\rho\alpha}^{\alpha} g^{\sigma\rho} )  - f \Delta_{\alpha\sigma}^{\sigma} \Delta_{\beta\rho}^{\rho} g^{\alpha\mu} g^{\beta\nu} \nonumber \\ +   \frac{f + f' \kappa}{2} \Delta_{\alpha\sigma}^{\sigma} \Delta_{\beta\rho}^{\rho}  g^{\alpha\beta} g^{\mu\nu}  + \frac{1}{2}  g^{\mu\nu} (e + e' k)  + \frac{16 \pi G}{\sqrt{-g} }  \frac{\delta S_{\rm mat}}{\delta g_{\mu\nu}} = 0. \end{eqnarray} One can split $S$ into $S_{e}$ and $S_{i}$ as before.  Employing the machinery used above in finding the generalized Bianchi identities and using the matter and gravitational equations of motion, one obtains \begin{eqnarray}   \frac{\delta S_{e}}{\delta \sqrt{-\eta} },_{\mu} = 0, \end{eqnarray} or, upon integration,  \begin{eqnarray} \frac{\delta S_{e}}{\delta \sqrt{-\eta} } = J, \end{eqnarray} where $J$ is a constant of integration.  The explicit form of $\frac{\delta S_{e}}{\delta \sqrt{-\eta} } $ is  \begin{eqnarray} \frac{\delta S_{e}}{\delta \sqrt{-\eta} } = \frac{\delta S_{\rm mat}}{\delta \sqrt{-\eta} } + \frac{1}{16 \pi G} ( - a' \kappa^2 R + \partial_{\nu}(2 f \nabla^{\nu}\kappa) - f' g^{\mu\nu} \kappa,_{\mu} \kappa,_{\nu} - e' \kappa^{2}).  \end{eqnarray}

   By making a conformal transformation to the Einstein frame, one can typically set $a = 1$.  One reason not to do so at this stage is because the above action contains degenerate cases related to \emph{unimodular} general relativity \cite{vanDam,UnruhUGR,Teitelboim,Dragon,Sorkin,Ng}, which involve $a = \sqrt{\kappa}$.  In these cases, the Ricci scalar term pertains to a curved metric whose determinant is just that of the flat metric and thus nondynamical; in searching for new theories, one wants not to lose sight of any special cases.  Also, nongravitational experiments are governed by the metric which is conformally coupled to matter (as will be discussed below), if one exists; typically that is not the Einstein frame's metric.  Otherwise, setting $a=1$ is convenient. 
 \section{Some Special Cases and Empirical Consequences}       Slightly bimetric theories split into a number of cases, among which are generalized Brans-Dicke (Bergmann-Wagoner \cite{Will}) theories, general relativity without or with a scalar field, unimodular general relativity  \cite{vanDam,UnruhUGR,Teitelboim,Dragon,Sorkin,Ng}, and some others.  General relativity itself is of course a trivial example of a slightly bimetric theory.  An attractive example of general relativity with a scalar field was briefly considered by Avakian and Grigorian \cite{Avakian}; however, their refutation of the theory, which corresponds to an unspecified  constant $a_{3}$ in their notation, cannot be accepted because the theory manifestly \emph{includes} general relativity, and thus every solution of the Einstein field equations, as a special case.  This theory is very similar to the ``restricted gravity'' of Dragon and Kreuzer, who find a massive dilaton in the metric \cite{DragonKreuzer}.  Unimodular general relativity sets $\sqrt{-g} = \sqrt{-\eta} $ \emph{a priori}, so the traceless part of the Einstein equations are the Euler-Lagrange equations.  The Bianchi identities restore the trace of the Einstein equations, up to an integration constant.  It is interesting to note that in considering the ``most general linear theory of gravitation'', Nachtmann, Schmidle, and Sexl omitted the case in which matter is coupled  only to the traceless part of the gravitational field \cite{NSSexlField,NSSexlLinear}.  Such a case corresponds to coupling to a covariantly unimodular matter metric in the nonlinear theory.   

    One readily sees that some slightly bimetric theories contain general relativity (perhaps with the covariantly unimodular condition $\kappa = 1$) as a special case.  Full consideration of the empirical properties of the theories requires dividing the family of theories into natural cases; the theories do not even all have the same number of degrees of freedom.  Various equivalence principles are satisfied, or violated, as the case may be, for particular slightly bimetric theories, so different versions might provide tests of various equivalence principles.  Theories in which matter is not universally coupled will tend to violate the weak equivalence principle \cite{ThorneLL}.   Because some slightly bimetric theories grade continuously into general relativity, these versions ought to remain viable as long as general relativity's outstanding track record persists.  Full consideration of these matters awaits another time. 

\subsection{Built-in Scalar Field?}       The scalar degree of freedom present in some slightly bimetric theories  could perhaps be detected once gravitational wave astronomy is well under way \cite{Fucito}.  In addition, it might facilitate inflationary cosmological models, because it can be nonminimally coupled, as inflation requires \cite{Veneziano}.  Or it might serve as a form of dark matter. There have been a number of studies of scalar field dark matter recently \cite{Matos}.  For minimally coupled matter, the scalar field acts as ``noninteracting dark matter,'' which interacts only with itself and gravity.  This form of dark matter has recently been considered by Peebles and Vilenkin \cite{PeeblesVilenkin}.       

     Using the scalar-tensor twin of a slightly bimetric theory should permit carrying over many results from scalar-tensor theories to slightly bimetric theories, such as issues of positive scalar field energy \cite{Santiago}.  

\subsection{Cosmological Constant Problem} 	

Concerning the cosmological constant, theorists have been interested in explaining the difference between its quantum-mechanically predicted large value and its observed small value---this is the ``cosmological constant problem'' \cite{Weinberg CC}. (At least, this is the ``old cosmological constant problem''; recently new cosmological constant problems have arisen \cite{Ng}.)  One approach that has attracted attention is unimodular general relativity \cite{vanDam,UnruhUGR,Teitelboim,Dragon,Ng}, because the cosmological constant is in that case not a coupling constant in the action, but a datum in the initial conditions.  Other slightly bimetric theories behave in the same fashion, the integration constant $J$ being related to an effective cosmological constant, so they retain this advantage in addressing this problem.  From a classical experimental point of view, it is thought to be necessary to include an effective cosmological constant.  Receiving it as a constant of integration is much more appealing than the traditional way by putting a term linear in the gravitational field into the action, for such an action defines a theory in which the field about a point source grows with distance, behavior which is difficult to accept \cite{Freund}.

   We note in passing that other authors have also modified the nature of the scalar densities in the action \cite{Guendelman}, albeit differently, with solving the cosmological constant problem in view.

\section{Interpretation of Bimetric Theories}

  \subsection{Generally Covariant Bimetric Theories}  

    It might be useful to explain why the bimetric/field approach to general relativity is empirically equivalent to the geometrical  form, at least in their classical regimes.\footnote{However, the bimetric theory's topology is restricted to be $R^{4}$ (or at least to be compatible with a flat metric).  But this limitation might be less strict than it seems, for it has been suggested that  spatially closed worlds can be accommodated using a flat topologically trivial background metric \cite{Zel1,GrishTopol}.  It is worth noting that these authors regard the flat metric as useful but fictitious, based on its unobservability \cite{Zel1} and the possibility that the curved metric's light cone might open wider than the flat metric's \cite{Zel2,Grishchuk90}, as will be discussed below.  For another view, see (\cite{VlasovTopol}).}  Questions might arise due to the fact that measurements of times and lengths in the geometrical  theory are \emph{assumed} to be governed by $g_{\mu\nu}$, there being no other metric tensor to choose; but if $\eta_{\mu\nu}$ is also present, then other choices might seem possible.  This proof will also help to give the empirical interpretation of slightly bimetric theories.

  If one considers what an `ideal' rod or clock might be, the geometrical view says that it is one governed by $g_{\mu\nu}$ \cite{Misner}, whereas the bimetric approach says that it is one that is unaffected by gravity and thus governed by $\eta_{\mu\nu}$.  But it is real rods and clocks, not ideal ones, that are used in experiments.  J. L. Anderson has recently argued that a metric in general relativity is unnecessary, because the behavior of rods and clocks can be determined \emph{via} the Einstein-Infeld-Hoffmann procedure \cite{AndersonMetric}.  Even if such a procedure were impossible in practice, it would remain true that the behavior of real rods and clocks would be completely determined (classically) by the partial differential equations obeyed by all the fields, for, in light of modern field theory, real rods and clocks are just congealed field excitations.  Conceptually, there is no room for a separate postulate of the behavior of length and time measurements.  Because the bimetric and geometrical approaches to general relativity yield identical partial differential equations for $g_{\mu\nu}$ and matter fields $u$, it follows that the two approaches are empirically equivalent.  Thus, once the obsolete dualism between matter and field is removed,\footnote{A quantum mechanical analog of our reasoning would be the insistence that measurement is not ultimately different from time evolution (with a sudden collapse of the wave function), but is only a particular case of evolution.} it becomes clear that these two approaches to general relativity are equivalent empirically, at least locally and classically.  This issue has also been addressed by Thirring \cite{Thirring}, Freund \emph{et al.} \cite{Freund}, and Zel'dovich and Grishchuk \cite{Zel1}.

 \subsection{Interpretation of Slightly Bimetric Theories}

   In the case of slightly bimetric theories, it is no longer the case that the flat background metric is entirely clothed.  So how does one interpret measurements?  Here the existence of a scalar-tensor ``twin'' for each slightly bimetric theory is useful. Assuming that the usual postulated relation between measurements in general relativity and the partial differential equations of general relativity is consistent, the same results can be carried over to slightly bimetric theories \emph{via} their scalar-tensor twins.  Scalar-tensor theories are specific examples of general relativity coupled to a scalar field.  In some theories, there exists a ``Jordan frame'' in which matter is minimally coupled, as in general relativity.  General relativity assumes nongravitational experiments to be described by the metric minimally coupled to matter. The scalar field should not make any difference, for one could regard it as a peculiar matter field.  So the relevant metric for typical experiments is the one minimally coupled to matter, if such a thing exists.

 \subsection{Tetrad Field and a Flat Metric} 

	A few comments on a tetrad field in general relativity are in order.  Concerning localization of gravitational energy-momentum, C. M{\o}ller concluded that a satisfactory solution within Riemannian geometry does not exist, but that one does exist in a tetrad form of general relativity, apart from the question of finding the `extra' 6 equations to fix the freedom under local Lorentz transformations \cite{Moller,MollerItaly,MollerWarsaw,MollerSurvey}. Some recent improvements in locally positive energy by Nester \emph{et al.} also make use of a tetrad field; see (\cite{NesterTele,Nester}) and references therein. The bimetric and tetrad formalisms are not unrelated \cite{OP,OPspinor,Davis}.   	

One interesting but little-noted connection between the two formalisms was found by Ogievetski\u{i} and Polubarinov \cite{OPspinor}.  They were able to find a substitute for a tetrad field in coupling fermions to gravitation.  They replace the tetrad with a `square root of the (curved) metric tensor' written explicitly as an infinite binomial series in the gravitational field (in this case equal to $\frac{ g^{\mu\nu} - \eta^{\mu\nu} }{-\lambda }$), along with the flat metric tensor.\footnote{Actually they use the matrix $diag(1,1,1,1)$ along with an imaginary time coordinate.  It is plain, however, how such work would generalize to the use of a real time coordinate with $diag(-1,1,1,1),$ and then from this matrix to a flat  metric \emph{tensor} $\eta_{\mu\nu}.$ Use of the latter permits taking a tensorial square root; this square root transforms nonlinearly under (non-coordinate) gauge transformations, however. The unmodified Ogievetski\u{i}-Polubarinov square root of the metric is a nonlinear geometric  object with respect to coordinate transformations. }  This quantity is symmetric and an ordinary tensor, as opposed to an asymmetric quantity with one vector index and one local Lorentz index.  Thus, it enjoys the simplicity of having only one sort of index and only 10 independent components.  We can envision several interesting consequences of using this quantity.  First, given a flat metric tensor, the existence of this quantity refutes the conventional claim \cite{VisserWorm} (p. 373) that a tetrad field is more fundamental than a symmetric tensor gravitational field. This fact might affect one's efforts at quantization (\emph{c. f.} \cite{VisserWorm,DeserVierbein}).  Second, one avoids the complexity of introducing extra variables and consequently many more constraints (\emph{c.f.} \cite{Charap}).  Third, by taking the ``square root'' quantity as the basic variable, one could use an \emph{a priori} symmetric ``tetrad'' in the tetrad formulation of general relativity. Thus, there would be no need to search for another six equations to fix the `extra' tetrad components.  (Alternatively, one might impose symmetry \emph{a posteriori} as a gauge condition.)  However, one possible difficulty  with \emph{a priori} symmetry is that one loses the freedom to choose the ``time gauge'' by attuning the temporal part of the tetrad to the time coordinate.

 \section{Dueling Null Cones?}       One important question concerning the acceptability of the field form of general relativity (and similar theories) involves the relation between the curved and flat metrics' null cones.  As was briefly mentioned earlier, if the special relativistic nature of the theory is to be taken seriously, then nothing may propagate outside the null cone of $\eta_{\mu\nu}$, on pain of causality violation.  Yet it is $g_{\mu\nu}$ that governs physical propagation.  Thus, as Penrose \cite{Penrose} (and Bi\v{c}\'{a}k following him \cite{Bicak}), Zel'dovich and Grishchuk \cite{Zel2,Grishchuk90}, Burlankov \cite{BurlankovCause} and Logunov \emph{et al.} \cite{LogunovFund,Chugreev,Pinson} have noted, consistency imposes the nontrivial requirement that the light cone for $g_{\mu\nu}$ everywhere be no wider than that for $\eta_{\mu\nu}$.    

	This fundamental issue has received less attention than one might expect, given the number of papers written from a flat spacetime viewpoint.  It has been mentioned in connection with the covariant perturbation approach to quantum gravity \cite{Wald}, but apparently not addressed fully.  Concerning this question (and another that we do not discuss), van Nieuwenhuizen explained that the particle physicists approach is to ignore it and hope that it goes away.\footnote{He wrote  \cite{van N cones,Rovelli}: \begin{quote} \ldots According to the particle physics approach, gravitons are treated on exactly the same basis as other particles such as photons and electrons.  In particular, particles (including gravitons) are always in flat Minkowski space and move \underline{as if} they followed their geodesics in curved spacetime because of the dynamics of multiple graviton exchange \ldots Pure relativists often become somewhat uneasy at this point because \ldots  one must decide before quantization which points are spacelike separated and which are timelike separated \ldots  However, it is only after quantization that the fully quantized metric field can tell us this spacetime structure \ldots  The strategy of particle physicists has been to ignore [this problem] for the time being, in the hope that [it] will ultimately be resolved in the final theory.\end{quote}} While quantization is not our present concern, the situation is similar at the classical level:  there is no obvious reason that the dynamics will yield a physical causal structure consistent with the \emph{a priori} special-relativistic one. The authors who have addressed the problem take several different stances on the subject.    

	Penrose and Bi\v{c}\'{a}k find a substantial objection to the field formulation, because Penrose shows that either the flat metric's null cone structure is violated, or the null geodesics of the two metrics diverge arbitrarily, far from any sources.  These two horns correspond to different gauge conditions.  Clearly the first horn is unsatisfactory.  However, we find that the latter problem can be traced merely to the long-range $\frac{1}{r}$ character of the potential in the conformally invariant part of the curved metric.  If the fall-off were a power law of the form $\frac{1}{r^{1 + \epsilon} }$, $\epsilon > 0$, then no difficulty would arise.  It is well-known that $\frac{1}{r}$ potentials have peculiar long-range scattering properties \cite{Goldstein}.  So the alleged difficulty follows immediately from the fact that a long-range spin-2 field is present. Penrose's objection to the second horn not being fatal, one can merely accept the second horn.  If a solution is needed, then adding a mass term suffices, at least if massive gravity can escape the traditional negative energy objection \cite{DeserMass} (appendix on ``ghost'' theories).  As we noted above, Visser has suggested that it can \cite{Visser}.  

  	Zel'dovich and Grishchuk are also confident that the light cone problem shows the flat metric to be fictitious, though they consider the fiction a useful one (see also (\cite{Babak})).  But their arguments fall short of a proof, largely because the conclusion strongly depends upon their specific gauge choice. Given their purpose in writing, they selected the gauge employed by Logunov \emph{et al.}, but one cannot assume that like results would obtain in all other gauges, especially in light of our argument below. 

   	Burlankov's position \cite{BurlankovCause} is fairly similar to that of Zel'dovich and Grishchuk, but a few points deserve special notice.  Burlankov is sympathetic to idea (asserted by Logunov \emph{et al.}) that general relativity has difficulties, noting ``the collapse problem, the singularity problem, strong gauge invariance, and the absence of a `natural' energy-momentum complex'' (p. 176).  However, Burlankov finds that the ``solution of the amazing problems in gravity does not lie'' in the bimetric formalism (p. 177).  And Minkowski space cannot be taken as fundamental.  Why not?  The difficulty is with the null cones.  However, we cannot agree with Burlankov that a consistent relation between the two light cones requires that the metrics be conformally related (p. 176), for it seems sufficient that the curved metric's null cone lie on \emph{or within} the flat cone.    

	Logunov \emph{et al.}, being committed to the flat spacetime view, see the question of compatible null cones as merely a problem to be solved, rather than a fatal flaw.  (We consider here only the older massless version of their theory, which uses the Einstein field equations and the tensorial DeDonder gauge.) Furthermore, they appear to believe it to be solved already by their own formulation.  They have set forth a causality principle, which we shall call the Logunov Causality Principle (LCP), that states that field configurations that make the curved metric's null cone open wider than the flat metric's are physically meaningless \cite{LogunovFund,Chugreev,Pinson}.  As they observe, satisfaction is not guaranteed (even with their gauge conditions, notes Grishchuk \cite{Grishchuk90}), which means that the set of partial differential equations is not enough to define the theory.  Some causality principle is indeed needed, but the LCP strikes us as somewhat arbitrary and \emph{ad hoc}.  One would desire three improvements.  First, one would prefer that the causality principle be closely tied to the equations of motion, not separately appended.  Second, one wants a guarantee that there exist enough solutions obeying the principle to cover all physically relevant situations.  Expressing the principle as a set of conditions on initial data and investigating their dynamical preservation might be a step in this direction. The fact that the conditions consist of inequalities, rather than equations, is not helpful. However, some mathematically analogous work has been done by Goldberg and Klotz in canonical general relativity \cite{GoldbergKlotz} (although we are interested in loose inequalities, while they employed strict inequalities).  Third, one would prefer a more convenient set of variables to describe the physics.  We hope to address these matters thoroughly in the future.

	For now we merely point out that Logunov's 4-dimensional analysis of the causality principle can be written surprisingly neatly using an ADM split \cite{Misner,Wald}.  Given the utility of such a split in the Hamiltonian form of general relativity \cite{Wald} and its massive relatives \cite{DeserMass}, this form might prove useful.  In considering whether all the vectors $V^{\mu}$ lying on $\eta$'s null cone are timelike, null, or spacelike with respect to $g$, it suffices to consider future-pointing vectors with unit time component; thus $V^{\mu} = (1, V^{i})$, where $V^{i} V^{i} = 1 $ (the sum running from 1 to 3) when Cartesian coordinates are used.  Using the $-+++$ signature, the causality principle can be written $h_{ij} (\beta^{i} + V^{i}) (\beta^{j} + V^{j}) - N^{2} \stackrel{?}{\geq} 0$ for all spatial unit vectors $V^{i}$.  Here the spatial metric is $h_{ij}$, the lapse is $N$, and the shift is $\beta^{i}$. 	 	

It is worth reiterating that the local relation between the light cones is gauge-dependent \cite{Penrose,Grishchuk90} in gauge-invariant theories, as we saw above with Penrose's dilemma.  This fact proves  that the gauge invariance needs to be broken (at least in part) in some suitable natural way, perhaps by adding a mass term or Lagrange mulipliers \cite{KucharHarmonic}
A satisfactory causality principle would judge an entire theory (including any gauge fixing and positivity conditions), not merely individual solutions, as physically acceptable or not, \emph{pace} Logunov \emph{et al}.

	A plausibility argument will now show that gauges satisfying the causality principle likely \emph{do} exist.  Given a flat background metric and a Cartesian coordinate system for it,  one can readily draw the flat and curved metrics' light cones on the tangent space at some event (apart from obvious difficulties with 4-dimensional pictures). One wants the curved cone to be located on or within the flat one.  (The flat cone has the usual ideal conical shape, whereas the curved one is distorted, in general.)  In a bimetric context, it is basically the case that the curved spatial metric controls the width of the light cone, while the shift vector determines its tilt from the vertical (future) direction and the lapse function determines its length.  For generally covariant theories such as general relativity, the spatial metric contains the physical degrees of freedom; the lapse and shift represent the gauge freedom, so they can be chosen arbitrarily, at least over some region. (For slightly bimetric theories, one has one fewer arbitrary function to choose.) A suitable gauge would preserve the proper relation between the light cones, given that it existed at some initial moment.  By analogy with conditions typically imposed in geometrical general relativity to avoid causality difficulties \cite{Wald}, one would prefer, if possible, that the curved light cone be strictly inside the flat light cone (\emph{i.e.}, be $\eta$-timelike), not tangent to it, because tangency indicates that the field is on the verge of (special relativistic) causality violation.  Under quantization, one might expect fluctuations to push the borderline case into the unacceptable realm, so it seems best to provide a cushion to avoid the problem, if possible.  But that might not be possible, if flat spacetime is to be a solution of the theory.   

	Let the desired relation between the null cones hold at some initial moment.  Also let the curved spatial metric and shift be such at some event in that moment that they tend to make the curved cone violate the flat one a bit later. By suitably reducing the lapse, one can lengthen the curved cone until it once again is safely inside the flat cone.  By so choosing the lapse at all times and places, one should be able to satisfy the causality principle at every event, if no global difficulties arise. (One can imagine that the Schwarzschild radius will require careful attention.) Implementing this procedure in an attractive and principled way is a further challenge.    

	We have written the previous paragraph as if the shift and spatial metric were fixed physically, while only the lapse is gauge freedom.  But because in fact both the lapse and the shift are gauge freedom, it follows that both the length and the tilt of the curved light cone relative to the flat are at our disposal.  Therefore it is all the more likely that any solution of physical interest can be expressed in a gauge obeying the causality principle.  For slightly bimetric theories, the picture is slightly less rosy, because only 3 arbitrary functions worth of gauge freedom exist.  But if the lapse is chosen to be one of those three, then the situation appears satisfactory, because the lapse alone can do the job.  

	In keeping with the need to show that enough solutions exist to cover all physically interesting cases, it will be useful to note that the causality principle is not in obvious conflict with some of the usual cosmological models of general relativity. Because the spatially flat Robertson-Walker models have conformally flat spacetime metrics \cite{Wald}, it is plain that there exists a gauge in which the curved metric's null cone is identical to the flat metric's.  One can find such a gauge by declaring that the coordinates that make explicit the curved metric's conformal flatness, are Cartesian with respect to the flat metric.  The same move works, at least in a large neighborhood, for the other Robertson-Walker metrics, which are also conformally flat \cite{InfeldSchild}.

	One expects that the causality principle itself will help to dictate the gauge conditions in general.  If it can be shown that general relativity or some similar theory satisfies the causality principle (with a suitable generic principled gauge choice, \emph{etc.}, as needed--as opposed to the present level of development, in which we choose the gauge \emph{ad hoc} by hand), then the  flat spacetime field version of gravity will rest on a firmer footing, and will be much more appealing than if the flat metric is merely a convenient fiction.   

 \section{Acknowledgments}  One of us (J. B. P.) was supported in part by the Robert A. Welch Foundation, grant no. F-0365, and thanks Drs. L. Shepley, M. Choptuik, S. Deser,   B. DeWitt, S. Weinberg, and V. A. Petrov and Mr. Ioannis Gkigkitzis for  helpful discussions or correspondence. 


\end{document}